\documentclass[10pt,conference]{IEEEtran}
\IEEEoverridecommandlockouts

\usepackage{cite}
\usepackage{amsmath,amssymb,amsfonts}
\usepackage{algorithmic}
\usepackage{graphicx}
\usepackage{textcomp}
\usepackage{xcolor}
\usepackage{booktabs}
\usepackage{multirow}
\usepackage{array}
\usepackage[hidelinks]{hyperref}
\usepackage{xspace}
\usepackage{hyperref}

\usepackage[most]{tcolorbox}
\usepackage[table]{xcolor}

\usepackage{arydshln} 

\usepackage{graphicx}

\newcommand{\modelname}{AdaRankLLM\xspace} 

\begin{document}

\title{
Rethinking the Necessity of Adaptive Retrieval-Augmented Generation\\
through the Lens of Adaptive Listwise Ranking
}

\author{
 \textbf{Jun Feng\textsuperscript{1}},
 \textbf{Jiahui Tang\textsuperscript{2}}, 
 \textbf{Zhicheng He\textsuperscript{2}},
 \textbf{Hang Lv\textsuperscript{2}},
 \textbf{Hongchao Gu\textsuperscript{2}},\\
 \textbf{Hao Wang\textsuperscript{2}},
\textbf{Xuezhi Yang\textsuperscript{1}},
 \textbf{Shuai Fang\textsuperscript{1}}
 \\
 \textsuperscript{1}Hefei University of Technology,
 \textsuperscript{2}University of Science and Technology of China
}

\maketitle

\begin{abstract}

Adaptive Retrieval-Augmented Generation aims to mitigate the interference of extraneous noise by dynamically determining the necessity of retrieving supplementary passages. However, as Large Language Models  evolve with increasing robustness to noise, the necessity of adaptive retrieval warrants re-evaluation.
In this paper, we rethink this necessity and propose AdaRankLLM, a novel adaptive retrieval framework. To effectively verify the necessity of adaptive listwise reranking, we first develop an adaptive ranker employing a zero-shot prompt with a passage dropout mechanism, and compare its generation outcomes against static fixed-depth retrieval strategies. Furthermore, to endow smaller open-source LLMs with this precise listwise ranking and adaptive filtering capability, we introduce a two-stage progressive distillation paradigm enhanced by data sampling and augmentation techniques.
Extensive experiments across three datasets and eight LLMs demonstrate that AdaRankLLM consistently achieves optimal performance in most scenarios with significantly reduced context overhead. Crucially, our analysis reveals a role shift in adaptive retrieval: it functions as a critical noise filter for weaker models to overcome their limitations, while serving as a cost-effective efficiency optimizer for stronger reasoning models. Our code is available at \href{https://github.com/USTC-StarTeam/adaptive-listwise-ranking-rag}{USTC-StarTeam/adaptive-listwise-ranking-rag}.

\end{abstract}


\section{Introduction}

\begin{figure*}
    \centering
    \includegraphics[width=0.96\textwidth,height=0.44\textwidth]{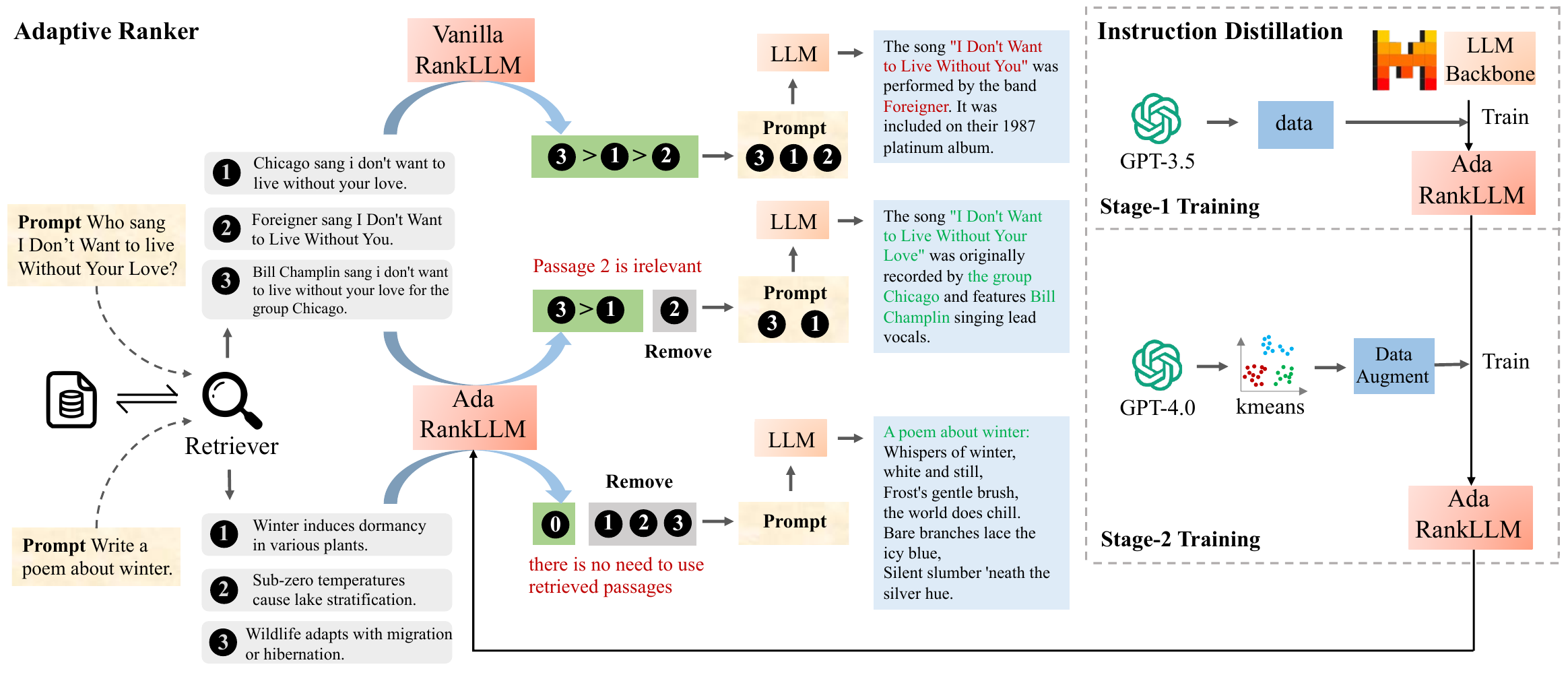}
    \caption{The framework of \modelname. The left part shows two examples to demonstrate how the Adaptive Ranker works.  The right part outlines the process of data collection and model training in order to distill an efficient ranker.}
    \label{fig: framework}
\end{figure*}

Retrieval-Augmented Generation (RAG) stands as a pivotal paradigm that augments Large Language Models (LLMs) by incorporating external knowledge~\cite{lewis2020retrieval,ma-etal-2023-query,lv2026costeercollaborativedecodingtimepersonalization,huang2025selfaugmitigatingcatastrophicforgetting}. By grounding generation in retrieved passages, RAG effectively mitigates hallucinations~\cite{zhang2023sirens} and facilitates continuous knowledge updates~\cite{ke2022continual}. However, its efficacy is not guaranteed; recent studies~\cite{ren2023investigating,shi2023large} show that indiscriminate retrieval often introduces extraneous noise. While standard RAG systems typically retrieve a fixed number of passages, this static strategy struggles to balance information coverage against noise intrusion: retrieving too few passages risks missing relevant information, while retrieving too many introduces noise that degrades model performance.

To address this dilemma, adaptive retrieval~\cite{asai2023self} has emerged as a dominant solution. By dynamically determining when and what to retrieve, it aims to shield models from irrelevant contexts and has historically yielded significant gains. However, the landscape of LLMs is rapidly evolving. Recent work~\cite{tu2025rbftrobustfinetuningretrievalaugmented} reveals that modern foundation models are becoming increasingly robust to retrieval noise, capable of ignoring irrelevant information via strong internal attention mechanisms. This evolution raises a fundamental question: \textit{Is adaptive retrieval still essential for these increasingly capable RAG systems?}

To investigate this question, we propose \modelname, a framework that synergizes Adaptive Ranking with Instruction Distillation. Our approach hinges on a rigorous mechanism to determine retrieval necessity: an adaptive ranker equipped with a Passage Dropout mechanism. Utilizing a zero-shot listwise prompt, this module evaluates passage relevance, explicitly discarding irrelevant candidates or outputting a termination token when no passages are relevant. To ensure this verification capability is not confined to cost-prohibitive large models, we further integrate a two-stage progressive distillation paradigm. By leveraging clustering-based sampling and data augmentation, we transfer listwise ranking and adaptive filtering skills from teacher models to smaller open-source backbones, allowing us to verify the utility of adaptive retrieval across varying model capabilities and deployment constraints.

We conduct extensive experiments on three benchmarks across eight LLMs, spanning models from smaller open-source backbones to advanced reasoning-enhanced architectures. The results demonstrate that \modelname consistently achieves optimal performance in most scenarios while significantly reducing context overhead. Crucially, our analysis reveals a clear distinction: for weaker models, adaptive retrieval acts as a critical \textit{noise filter} to ensure generation quality; for advanced models, while it offers limited performance gains, it serves as a vital \textit{efficiency optimizer} by significantly reducing inference costs.

Our main contributions can be summarized as follows:
\begin{itemize}
\item We examine the necessity of adaptive RAG for modern LLMs, moving beyond the traditional view to evaluate its distinct utility across models with varying capabilities.
\item We propose \modelname, a framework that employs an adaptive ranker with a passage dropout mechanism and instruction distillation to endow smaller models with precise noise-filtering capabilities.
\item Extensive experiments across eight LLMs reveal a functional distinction: adaptive retrieval acts as a critical noise filter for weaker models to ensure quality, while serving as a vital efficiency optimizer for advanced models.
\end{itemize}

\newpage

\section{Related Work}

\noindent\textbf{Passage Ranking.} 
Traditional ranking methods mainly rely on pointwise~\cite{sachan2022improving} or pairwise~\cite{qin2023large} scorers, yet they often overlook passage interrelations, prompting the shift to LLM-based listwise approaches~\cite{ma2023zeroshot,liu2024leveraging,lv2026learningemptinessdebiasinglistwise}. While early distillation attempts like RankVicuna~\cite{pradeep2023rankvicuna} and RankZephyr~\cite{pradeep2023rankzephyr} successfully transferred listwise capabilities to 7B models, they remained constrained by the quadratic complexity of decoder-only architectures. Recent advancements address these limitations: LiT5~\cite{tamber2025lit} reintroduces the Fusion-in-Decoder (FiD) architecture for linear scaling, while Rank-R1~\cite{zhuang2025setwise} integrates Chain-of-Thought (CoT) reasoning via Group Relative Policy Optimization (GRPO) to explicitly justify relevance. Further reducing latency, non-generative approaches utilize passage embeddings~\cite{mu2024perank} or attention patterns~\cite{icr2025} directly. Our method aligns with this trajectory, enhancing zero-shot listwise distillation with adaptive capabilities.


\noindent\textbf{Adaptive Retrieval.} 
Recent studies~\cite{ren2023investigating,shi2023large,jeong2024adaptive,liu2024ctrla} indicate that RAG performance is easily compromised by extraneous information. Early approaches like FLARE~\cite{jiang-etal-2023-active} and Self-RAG~\cite{asai2023self} relied on token probabilities or learned reflection tokens, often resulting in high inference or fine-tuning costs. Recent research pivots toward internal state monitoring: CTRLA~\cite{ctrla2025} and TARG~\cite{targ2025} leverage ``honesty'' directions and prefix logits to gate retrieval without auxiliary classifiers. For complex reasoning, FAIR-RAG~\cite{fairrag2025} , REPAIR~\cite{repair2026} and RAPID~\cite{gu2025rapid} introduce structured, plan-aware loops to trigger ``mid-course'' retrieval based on intermediate steps rather than the initial query. Additionally, cost-aware reinforcement learning strategies~\cite{li2025cost} dynamically adjust retrieval depth.
However, their complexity hampers zero-shot generalization and deployment efficiency, motivating our lightweight framework.

\section{Method}

\subsection{Problem Definition}
Given a user query $x$ and a set of candidate passages $P=\{p_1, p_2, ..., p_m\}$ retrieved from a massive corpus $\mathcal{C}$ (where $m$ is the number of candidates), the goal of Adaptive Retrieval-Augmented Generation is to identify an optimal ordered subset $A \subseteq P$. This process aims to maximize the relevance of the selected passages to $x$ while filtering out noise. Crucially, the size of the result set $|A|$ is dynamic and dependent on the query complexity. The set $A$ can range from the empty set $\emptyset$ (when retrieval is unnecessary) to the full candidate set. Finally, the selected passages $A$ and query $x$ serve as input to the generator to produce the response $y = \text{LLM}(A; x)$.

\subsection{Adaptive Ranker with Passage Dropout}
\label{sec:3/2}
To transform the coarse candidate set $P$ into the refined adaptive set $A$, we propose an \textbf{Adaptive Ranker} that advances the paradigm of listwise reranking. Unlike conventional rerankers that merely permute the fixed candidate list, our module integrates a passage dropout mechanism to dynamically filter extraneous noise while capturing the interrelations among passages.

\begin{figure}[t]
  \centering
  \includegraphics[width=0.95\linewidth,keepaspectratio]{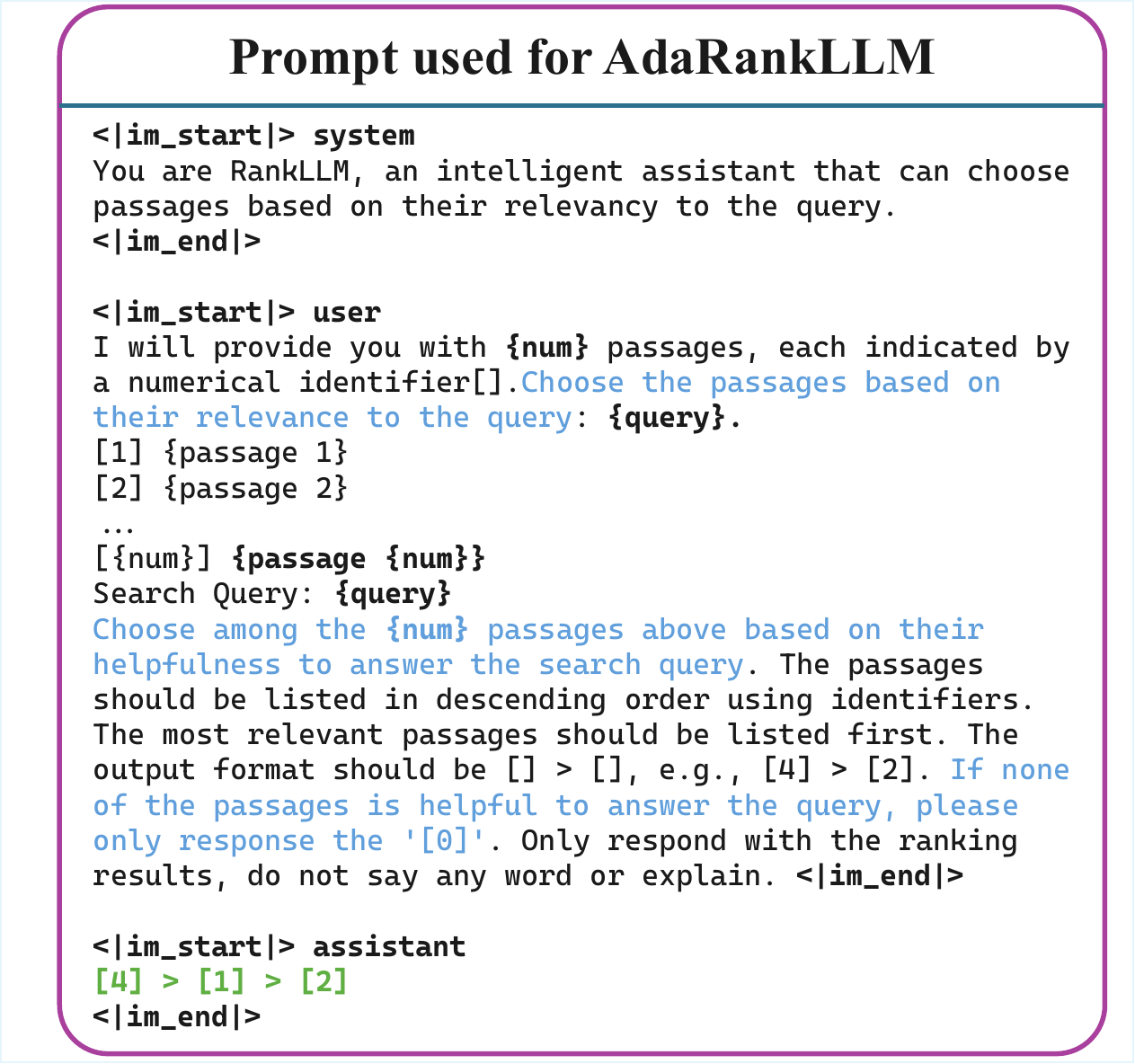}
  \vspace{-3mm}
\caption{Illustration of the prompt template used in \modelname for relevance-based passage selection and reranking.}
  \vspace{-3mm}
  \label{fig:prompt}
\end{figure}

Specifically, we design a zero-shot conversational prompt to instruct the LLM to perform relevance assessment and selection simultaneously. Each passage in the candidate set $P$ is assigned a unique identifier (e.g., `[1]'). Instead of forcing a full permutation, the LLM is directed to select only the passages that contribute to answering the query and rank them by relevance. Crucially, the passage dropout mechanism enables a rejection option: if the model determines that a passage (or even the entire set) is irrelevant, it explicitly excludes it from the output list $A$. In extreme cases where no relevant information is found, the model outputs a special termination token `[0]'. The specific prompt structure is illustrated in Figure~\ref{fig:prompt}. Following this adaptive ranking process, we parse the generated identifier list to construct the final context set $A=\{p_{r_1}, ..., p_{r_n}\}$, where the subset size $n$ is dynamically determined by the model's relevance judgment.

\subsection{Instruction Distillation}
While proprietary LLMs like GPT-4 demonstrate exceptional zero-shot adaptive ranking capabilities, their prohibitive costs hinder direct large-scale deployment. Consequently, we aim to distill these capabilities into a compact open-source model using a progressive two-stage paradigm to balance distillation cost and data quality.

\noindent\textbf{Stage 1: Structural Alignment.}
We utilize GPT-3.5 to generate a large-scale dataset from MSMARCO V1. This cost-effective initialization teaches the student model the fundamental listwise output schema and basic relevance ordering, establishing a solid structural foundation before.

\noindent\textbf{Stage 2: Adaptive Refinement.}
To inject precise passage dropout capabilities, we rely on GPT-4's superior reasoning. To minimize costs, we apply K-means clustering to query embeddings to sample a representative subset. GPT-4 annotates these selected queries to fine-tune the student's decision boundaries for distinguishing valid evidence from noise.

Finally, by integrating the adaptive ranker with passage dropout and instruction distillation, we will get an \modelname that is capable of selecting necessary passages, while maintaining efficiency and accessible.

\section{Experimental Setup}

\begin{table}[t!]
\caption{The three datasets employed in the experiments.} 
\label{tab:datasets}
\centering
{   \resizebox{\linewidth}{!}{
    \fontsize{8pt}{11pt}\selectfont
    \begin{tabular}{ccccc}
    \toprule
    \textbf{Datasets}  & \textbf{Corpus(\#passages)} & \textbf{Question Type} & \textbf{Retriever}\\
    \midrule
    ASQA  &Wikipedia(21M) & Factoid(ambiguous) & GTR\\
    QAMPARI  & Wikipedia(21M) & Factoid(list) & GTR\\
    ELI5  & Sphere(899M) & Why/How/What & BM25\\
    \bottomrule
    \end{tabular}

}
    
}
\end{table}

\subsection{Datasets}

We validate our method on three diverse public datasets: \textbf{ASQA}~\cite{stelmakh2022asqa}, \textbf{QAMPARI}~\cite{amouyal2023qampari}, and \textbf{ELI5}~\cite{fan2019eli5}. Consistent with standard protocols~\cite{gao2023enabling}, we employ Exact Match (EM) for ASQA, F1 score for QAMPARI, and Claim Recall for ELI5. To evaluate comprehensive capabilities, we also report an \textit{Overall*} metric computed as the average of these scores. Detailed dataset information are provided in Table~\ref{tab:datasets}.

\subsection{Baselines}

We compare \modelname with several baselines categorized into three groups:
(1) \textbf{Vanilla-$k$}: Models utilizing retrieved passages, where k denotes the number of passages provided to the LLM.
(2) \textbf{Rerank-$k$}: Models employing reranked passages, with k indicating the number of reranked passages supplied to the LLM. RankGPT4~\cite{sun2023chatgpt} is used as the reranker.
(3) \textbf{AdaRank-GPT4}: Models utilizing our proposed adaptive ranking prompt (Section \ref{sec:3/2}) with GPT-4, serving as a baseline to verify the necessity of adaptive RAG.
Additionally, we report the \textbf{Oracle} values for each backbone, representing the theoretical optimal performance when selecting the best number of reranked passages ($k\in[0,10]$).

\subsection{Models}

To comprehensively investigate the necessity of adaptive retrieval across varying model capabilities, we conduct experiments on seven backbone models: {Alpaca-7b}, Mistral-7b, GPT-3.5, GPT-4o, Llama-3.1-8B-Instruct, Qwen2.5-7B-Instruct, and Qwen3-8B (specifically evaluated with \texttt{enable\_thinking=False} and \texttt{True}). These models span a diverse spectrum of open-source and proprietary architectures.

\subsection{Implementation Details}

\paragraph{Training Details}
We fine-tune \modelname on Mistral-7B-v0.1~\cite{jiang2023mistral} using a subset of the MSMARCO V1~\cite{nguyen2016ms} dataset. The data comprises 100K queries with the top-20 BM25-retrieved passages, truncated to 10 to align with typical RAG settings. We use all 100K queries for the first stage and sample 5K queries for the second stage using \textbf{ADA2} embeddings~\cite{neelakantan2022text} with k-means clustering ($k=20$). Training runs for three epochs per stage with a learning rate of $5\text{e-}6$ (cosine decay) and a batch size of 64. 

\paragraph{Evaluation Details}
For all three datasets, We utilize the pre-processed first-stage retrieval results from ALCE~\cite{gao2023enabling}\footnote{\url{https://huggingface.co/datasets/princeton-nlp/ALCE-data}} and apply official chat templates for all backbone models. Regarding metric implementation, we strictly follow the ALCE codebase: employing STR-EM for ASQA and the T5-11B based TRUE model~\cite{honovich-etal-2022-true-evaluating} to verify claim entailment for ELI5. 

\begin{table*}[t!]
\centering
\footnotesize
\renewcommand{\arraystretch}{1.05}
\setlength{\tabcolsep}{3.5pt}

\caption{Results for Alpaca-7B, Mistral-Instruct, GPT-3.5, GPT-4o, Llama, Qwen2.5-7B, and Qwen3-8B. For thinking model Qwen3-8B, we explicitly evaluate two configurations: \texttt{enable\_thinking=False} and \texttt{enable\_thinking=True}. The best results are highlighted in \textbf{bold}, and the second-best results are \underline{underlined}.}
\label{tab:main_result_gray}

\resizebox{0.9\textwidth}{!}{%
\begin{tabular}{l c c c c c c c c c c c c}
\toprule
\multirow{2}{*}{Method} & \multicolumn{1}{c}{\bf ASQA} & \multicolumn{1}{c}{\bf ELI5} & \multicolumn{3}{c}{\bf QAMPARI} & & \multicolumn{1}{c}{\bf ASQA} & \multicolumn{1}{c}{\bf ELI5} & \multicolumn{3}{c}{\bf QAMPARI} & \\
\cmidrule(lr){2-2} \cmidrule(lr){3-3} \cmidrule(lr){4-6} \cmidrule(lr){8-8} \cmidrule(lr){9-9} \cmidrule(lr){10-12}
 & \bf EM & \bf Claim & \bf Prec. & \bf Rec. & \bf F1 & \bf Overall$^*$ & \bf EM & \bf Claim & \bf Prec. & \bf Rec. & \bf F1 & \bf Overall$^*$ \\

\midrule
 & \multicolumn{6}{c}{\textbf{Alpaca-7B}} & \multicolumn{6}{c}{\textbf{Mistral-Instruct}} \\
\cmidrule(lr){2-7} \cmidrule(lr){8-13}
Vanilla-0   & 19.03 & 10.17 & 21.53 & 5.81 & 7.96 & 12.55 & 27.61 & \underline{15.80} & 12.87 & 13.79 & 11.58 & 18.33 \\
Vanilla-1   & 31.09 & 12.33 & 26.21 & 8.72 & 11.07 & 18.16 & 36.30 & 13.83 & 22.01 & 10.08 & 11.58 & 20.57 \\
Vanilla-3   & 32.70 & \underline{13.73} & 26.12 & 9.82 & 11.85 & 19.43 & 44.46 & 14.83 & 21.24 & 11.87 & 12.69 & 24.05 \\
Vanilla-5   & 32.75 & 13.33 & 25.21 & 10.03 & 11.93 & 19.34 & 46.23 & 14.70 & 21.66 & 14.52 & 14.27 & 25.07 \\
Vanilla-10  & 29.99 & 12.70 & 25.62 & 9.04 & 11.34 & 18.01 & 47.23 & 15.43 & 21.14 & 16.70 & 15.40 & 26.02 \\
\cdashline{1-13}
\addlinespace[2pt]
Rerank-1    & 35.01 & 13.10 & \textbf{31.95} & 10.40 & 13.19 & 20.43 & 43.17 & 13.83 & \textbf{28.26} & 12.09 & 13.71 & 23.57 \\
Rerank-3    & \underline{35.03} & \textbf{14.13} & 29.00 & \underline{11.28} & \underline{13.57} & \textbf{20.91} & 47.36 & 14.83 & \underline{25.84} & 14.66 & 15.52 & 25.90 \\
Rerank-5    & 33.85 & 13.43 & 27.76 & 10.95 & 12.97 & 20.08 & \underline{47.55} & 14.70 & 23.24 & 16.20 & 15.96 & 26.07 \\
Rerank-10   & 29.51 & 12.57 & 24.99 & 9.20 & 11.41 & 17.83 & 46.81 & 15.43 & 20.19 & 16.78 & 15.21 & 25.82 \\
\cdashline{1-13}
\addlinespace[2pt]
\rowcolor{gray!15} AdaRankGPT4 & \textbf{35.22} & 12.93 & \underline{31.29} & \textbf{11.51} & \textbf{14.18} & \underline{20.78} & 46.66 & \textbf{16.03} & 25.39 & \textbf{17.48} & \textbf{17.22} & \underline{26.64} \\
\rowcolor{gray!15} AdaRankLLM  & 33.86 & 13.07 & 28.70 & 11.06 & 13.53 & 20.15 & \textbf{48.14} & 15.23 & 24.05 & \underline{17.18} & \underline{16.77} & \textbf{26.71} \\
\cdashline{1-13}
\rowcolor{red!15} Oracle      & 46.78 & 25.77 & - & - & 27.49 & 33.35 & 64.53 & 28.93 & - & - & 33.23 & 42.23 \\

\midrule
 & \multicolumn{6}{c}{\textbf{GPT-3.5}} & \multicolumn{6}{c}{\textbf{GPT-4o}} \\
\cmidrule(lr){2-7} \cmidrule(lr){8-13}
Vanilla-0   & 41.86 & 18.63 & 28.47 & 24.48 & 24.22 & 28.24 & 45.25 & \textbf{27.33} & 32.35 & 30.02 & 30.92 & 34.50 \\
Vanilla-1   & 39.95 & 14.33 & 25.38 & 12.20 & 14.49 & 22.92 & 42.45 & 20.67 & 31.25 & 14.01 & 17.22 & 26.78 \\
Vanilla-3   & 45.44 & 16.67 & 29.17 & 10.85 & 19.84 & 27.32 & 49.85 & 20.33 & 31.47 & 19.40 & 22.15 & 30.78 \\
Vanilla-5   & 46.20 & \underline{18.67} & 30.23 & 21.66 & 22.40 & 29.09 & \textbf{53.38} & 22.33 & 35.87 & 24.20 & 26.82 & 34.18 \\
Vanilla-10  & \textbf{51.32} & 18.33 & 32.42 & \textbf{27.32} & 26.77 & \textbf{32.14} & 50.92 & 23.67 & 39.04 & 29.60 & 31.19 & 35.26 \\
\cdashline{1-13}
\addlinespace[2pt]
Rerank-1    & 41.38 & 16.33 & \textbf{39.30} & 16.40 & 20.38 & 26.03 & 44.55 & 22.33 & 39.94 & 18.01 & 22.06 & 29.65 \\
Rerank-3    & \underline{49.55} & 15.33 & 36.28 & 21.80 & 24.61 & 29.83 & 50.67 & 21.33 & 38.29 & 24.60 & 27.39 & 33.13 \\
Rerank-5    & 48.78 & 15.67 & 35.35 & 22.81 & 25.39 & 29.95 & 49.30 & 21.67 & 38.11 & 28.23 & 30.28 & 33.75 \\
Rerank-10   & 47.02 & 17.67 & 36.27 & \underline{26.82} & \textbf{28.22} & 30.97 & 48.63 & 21.67 & 36.17 & 29.81 & 30.54 & 33.62 \\
\cdashline{1-13}
\addlinespace[2pt]
\rowcolor{gray!15} AdaRankGPT4 & 46.97 & \textbf{19.03} & \underline{36.65} & 25.68 & 27.15 & \underline{31.04} & \underline{51.97} & 22.67 & \textbf{43.05} & \underline{31.80} & \underline{33.67} & \textbf{36.10} \\
\rowcolor{gray!15} AdaRankLLM  & 45.27 & 17.02 & 36.25 & 25.59 & \underline{27.25} & 29.84 & 49.05 & \underline{24.04} & \underline{41.02} & \textbf{32.60} & \textbf{33.74} & \underline{35.60} \\
\cdashline{1-13}
\rowcolor{red!15} Oracle      & 66.80 & 31.05 & - & - & 34.81 & 44.22 & 64.20 & 37.00 & - & - & 46.07 & 49.09 \\

\midrule
 & \multicolumn{6}{c}{\textbf{Llama3.1-8B-Instruct}} & \multicolumn{6}{c}{\textbf{Qwen2.5-7B-Instruct}} \\
\cmidrule(lr){2-7} \cmidrule(lr){8-13}
Vanilla-0   & 35.49 & \textbf{18.37} & 16.98 & 23.10 & 16.86 & 23.57 & 27.02 & \textbf{20.50} & 15.24 & 10.06 & 11.05 & 19.52 \\
Vanilla-1   & 32.93 & 11.83 & 22.63 & 12.00 & 13.69 & 19.48 & 35.83 & 14.90 & 24.06 & 10.70 & 13.18 & 21.30 \\
Vanilla-3   & 41.30 & 13.50 & 24.63 & 18.38 & 18.70 & 24.50 & 43.91 & 17.13 & 26.59 & 15.88 & 17.74 & 26.26 \\
Vanilla-5   & 44.75 & 15.13 & 25.45 & 23.56 & 21.77 & 27.22 & 45.58 & 17.00 & 28.72 & 18.46 & 19.85 & 27.48 \\
Vanilla-10  & 48.31 & 17.10 & 25.73 & \underline{28.56} & 23.73 & 29.71 & \underline{49.05} & 18.80 & 29.08 & 21.86 & 22.15 & 30.00 \\
\cdashline{1-13}
\addlinespace[2pt]
Rerank-1    & 38.28 & 13.34 & \textbf{30.52} & 15.94 & 18.23 & 23.28 & 42.49 & 16.47 & \textbf{33.08} & 14.26 & 17.53 & 25.50 \\
Rerank-3    & 44.57 & 15.30 & \underline{30.06} & 23.54 & 23.51 & 27.79 & 47.25 & 17.50 & \underline{31.59} & 19.20 & 21.06 & 28.60 \\
Rerank-5    & 47.04 & 15.67 & 28.41 & 26.52 & \underline{24.41} & 29.04 & \textbf{49.16} & 18.20 & 29.69 & 20.98 & 21.88 & 29.75 \\
Rerank-10   & \textbf{48.92} & 17.20 & 25.66 & \textbf{29.42} & 24.02 & \textbf{30.05} & \underline{49.05} & 18.47 & 30.31 & \underline{22.16} & \underline{22.58} & \underline{30.03} \\
\cdashline{1-13}
\addlinespace[2pt]
\rowcolor{gray!15} AdaRankGPT4 & 45.88 & 16.50 & 30.02 & 28.22 & \textbf{25.54} & 29.31 & 47.84 & \underline{19.77} & 31.19 & \textbf{22.30} & \textbf{22.84} & \textbf{30.15} \\
\rowcolor{gray!15} AdaRankLLM  & 47.44 & \underline{17.50} & 27.31 & 28.04 & 24.29 & \underline{29.74} & 47.40 & 18.60 & 29.81 & 21.76 & 22.26 & 29.42 \\
\cdashline{1-13}
\rowcolor{red!15} Oracle      & 63.32 & 30.47 & - & - & 37.52 & 43.77 & 60.50 & 32.33 & - & - & 31.93 & 41.58 \\

\midrule
 & \multicolumn{6}{c}{\textbf{Qwen3-8B-NoThinking}} & \multicolumn{6}{c}{\textbf{Qwen3-8B-Thinking}} \\
\cmidrule(lr){2-7} \cmidrule(lr){8-13}
Vanilla-0   & 26.79 & \textbf{20.67} & 14.32 & 7.98 & 13.65 & 20.37 & 28.93 & \textbf{23.27} & 13.69 & 8.05 & 13.37 & 21.85 \\
Vanilla-1   & 39.06 & 16.37 & 20.81 & 8.23 & 15.13 & 23.52 & 41.51 & 16.10 & 22.54 & 7.99 & 14.51 & 24.04 \\
Vanilla-3   & 47.51 & 17.93 & 23.79 & 11.36 & 19.94 & 28.46 & 48.93 & 17.60 & 25.91 & 12.12 & 20.49 & 29.01 \\
Vanilla-5   & 49.24 & 18.63 & 25.02 & 13.12 & 22.51 & 30.13 & 51.59 & 18.37 & 27.49 & 14.55 & 23.31 & 31.09 \\
Vanilla-10  & \textbf{52.77} & 20.17 & 26.30 & 15.78 & 25.21 & \textbf{32.72} & \textbf{54.85} & 21.03 & 28.50 & \textbf{17.61} & \textbf{26.66} & \underline{34.18} \\
\cdashline{1-13}
\addlinespace[2pt]
Rerank-1    & 44.42 & 17.57 & 27.37 & 10.08 & 18.99 & 26.99 & 46.00 & 17.93 & \textbf{31.05} & 10.46 & 18.95 & 27.63 \\
Rerank-3    & 48.97 & 18.53 & \textbf{29.04} & 14.50 & 25.08 & 30.86 & 50.63 & 19.57 & \underline{30.36} & 15.28 & 24.85 & 31.68 \\
Rerank-5    & 50.92 & 19.73 & \underline{28.27} & 15.42 & 25.74 & 32.13 & 52.64 & 19.10 & 29.22 & 16.57 & 25.81 & 32.52 \\
Rerank-10   & \underline{51.57} & \underline{20.33} & 26.68 & 16.27 & 25.51 & \underline{32.47} & \underline{54.82} & \underline{22.00} & 28.91 & \underline{17.58} & \underline{26.48} & \textbf{34.43} \\
\cdashline{1-13}
\addlinespace[2pt]
\rowcolor{gray!15} AdaRankGPT4 & 50.34 & 19.77 & 28.24 & \underline{16.30} & \textbf{26.23} & 32.11 & 52.17 & 21.90 & 28.66 & 17.54 & 26.20 & 33.42 \\
\rowcolor{gray!15} AdaRankLLM  & 50.40 & 19.93 & 28.16 & \textbf{16.40} & \underline{26.09} & 32.14 & 51.00 & 21.50 & 28.52 & 17.01 & 25.76 & 32.75 \\
\cdashline{1-13}
\rowcolor{red!15} Oracle      & 62.69 & 32.29 & - & - & 36.43 & 43.81 & 63.99 & 33.99 & - & - & 37.52 & 45.17 \\

\bottomrule
\end{tabular}%
}
\end{table*}

\clearpage

\clearpage
\section{Results}

\subsection{main result}


Table~\ref{tab:main_result_gray} presents the comparative performance across diverse models. Beyond demonstrating universal efficacy, the results underscore a critical \textbf{functional shift} in adaptive retrieval relative to model capability. While \modelname consistently approximates the optimal static strategies for each architecture, we observe that its primary utility evolves fundamentally: transitioning from an essential \textit{noise filter} for weaker models to a strategic \textit{efficiency optimizer} for stronger ones.

\noindent\textbf{For weaker models, AdaRankLLM acts as a critical noise filter.}
Models with limited parameter counts or reasoning capabilities (e.g., Alpaca-7b, Mistral-7b) exhibit low noise robustness, causing their performance to degrade when exposed to larger retrieval contexts (Vanilla-10 / Rerank-10). For these models, the external purification provided by our adaptive ranking is essential. \modelname effectively discards irrelevant passages that would otherwise distract the model, matching the high precision of aggressive filtering strategies while maintaining the flexibility to recall more information when necessary.

\noindent\textbf{For stronger models, AdaRankLLM serves as an efficiency optimizer.}
Advanced models, particularly those with explicit reasoning paths (e.g., Qwen3), possess strong internal attention mechanisms capable of handling noisy contexts (Vanilla-10) to maximize recall. However, this robustness comes at the cost of processing excessive irrelevant tokens. In these cases, \modelname maintains competitive performance—often indistinguishable from full-context retrieval—but drastically reduces the input length. This demonstrates that processing the full candidate set is often redundant; \modelname effectively extracts the ``minimal sufficient context'', eliminating the computational overhead of irrelevant tokens while preserving the critical evidence required for reasoning.

\subsection{Further Analysis}


\paragraph{\textbf{The Fallacy of Static Retrieval.}}
Our analysis demonstrates that the optimal retrieval depth ($k$) is highly volatile, shifting significantly depending on the specific task, dataset, and backbone model. This variance renders any universal fixed-$k$ configuration suboptimal—either risking information loss or incurring unnecessary computational costs. \modelname addresses this inherent unpredictability by dynamically calibrating context size for each query. It effectively eliminates the need for manual tuning, serving as an adaptive solution that autonomously navigates the trade-off between sufficiency and efficiency across diverse experimental setups.

\paragraph{\textbf{Efficacy of Instruction Distillation.}}
To validate our distillation paradigm, we compare the performance of our distilled model directly against the teacher model, AdaRank-GPT4. Results indicate that the student model achieves performance parity with GPT-4 across evaluations, effectively mastering the complex criteria for listwise ranking and adaptive passage dropout. This confirms that the high-cost reasoning capabilities of proprietary models can be successfully compressed into smaller, open-source architectures, making advanced adaptive retrieval viable for efficient local deployment.

\paragraph{\textbf{Noise Sensitivity of Weaker Architectures.}}
Weaker backbones (e.g., Alpaca, Mistral) exhibit high noise sensitivity, consistently achieving peak performance at minimal retrieval depths ($k=1, 3$). Constrained by limited reasoning capabilities and effective context windows, these models degrade rapidly when exposed to larger, noisier retrieval sets. Consequently, they necessitate aggressive noise filtration rather than broad ingestion. \modelname excels in this scenario by effectively ``purifying"" the context, automatically balancing precision and recall to align with the model's restricted processing capacity.


\paragraph{\textbf{Noise Robustness and Internal Verification in Advanced Models}} In stark contrast to weaker counterparts, advanced models favor maximum information intake, achieving peak performance under high-recall settings (e.g., Vanilla-10). This indicates that for capable architectures, \emph{information recall} takes precedence over precision. Utilizing sophisticated attention mechanisms, they isolate valid evidence from noise, meaning they suffer more from missing information due to aggressive filtering than from the presence of irrelevant data. This robustness is amplified in models with explicit \emph{Thinking} modes. We observe that Vanilla-10 often yields the best results here, suggesting that explicit reasoning~\cite{wei2022chain} acts as a potent \textbf{internal verification mechanism}. By logically distinguishing conflicting data step-by-step, these models ``internalize'' the noise-filtering step, enabling them to exploit the full breadth of raw retrieval sets without being misled by inherent noise.

\paragraph{\textbf{Performance Gap to Theoretical Optima}}Finally, across all evaluated architectures, a persistent disparity remains between the best achieved performance and the theoretical upper bound represented by the idealized ``Oracle''. This observation underscores that the fundamental challenges of Retrieval-Augmented Generation remain far from being fully resolved. The persistence of this gap suggests that strategies relying solely on increasing model scale, expanding context windows, or employing rudimentary adaptive filtering are beginning to yield diminishing marginal returns, highlighting the critical need for more sophisticated retrieval-integration paradigms to effectively bridge this disparity.

\section{Conclusion}
In this paper, we propose AdaRankLLM, a lightweight framework that endows smaller models with adaptive ranking capabilities via instruction distillation. Our experiments reveal that the primary utility of adaptive retrieval shifts with model capability: it functions as a critical noise filter for weaker models and an efficiency optimizer for stronger ones. Furthermore, the persistent gap between current state-of-the-art results and the theoretical ``Oracle"" indicates that despite the continuous advancement of model capabilities, the fundamental challenge of optimal information retrieval remains unresolved. Future work will investigate agentic paradigms to enable adaptive retrieval to approximate the theoretical optimal solution.

\bibliographystyle{IEEEtran}
\bibliography{custom}

\newpage
\appendix
\section{Broader Connections and Potential Extensions}
Although this paper focuses on open-domain question answering, the core idea behind AdaRankLLM---using listwise reasoning to adaptively retain only the minimally sufficient evidence---connects to a broader set of recent efforts on retrieval-augmented generation, generative reranking, and retrieval-aware decision making.

\subsection{Connections to Adaptive and Efficient RAG}
Recent studies suggest that retrieval is increasingly evolving from a static prepend-only module into a controllable component that must coordinate with downstream generation objectives. Beyond question answering, this trend also appears in schema-aware extraction, long-text generation, continual adaptation, and more challenging RAG evaluation settings \cite{liang-etal-2025-adaptive,gu2025rapidefficientretrievalaugmentedlong,huang2025selfaugmitigatingcatastrophicforgetting,zhang2025ragigbenchinnovativeevaluationragbased}. In this context, AdaRankLLM can be viewed as a lightweight alternative that studies how far adaptive evidence selection can go without introducing complex iterative control loops.

\subsection{Connections to Generative Ranking and Reranking}
AdaRankLLM is also related to the broader line of treating retrieval and ranking as generative decision problems. Recent works have explored content-agnostic calibration for generative listwise rerankers, semi-autoregressive personalized reranking with distillation, and unified formulations that couple retrieval and ranking \cite{lv2026learningemptinessdebiasinglistwise,cheng2026efficientpersonalizedrerankingsemiautoregressive,zhang2025killingbirdsstoneunifying,zhi2026spardselfpacedcurriculumrl}. Compared with these studies, our focus is not only to improve ranking quality over a fixed candidate set, but also to determine a variable-sized evidence set---including the null case---so that reranking becomes a mechanism for estimating retrieval necessity.

\subsection{Potential Extensions Beyond QA}
The principle of adaptive evidence pruning may also be relevant beyond QA, especially in settings where over-retrieval is costly, distracting, or privacy-sensitive. Examples include personalized generation, edge-cloud collaboration, user-centric agents, retrieval-enhanced recommendation, and interactive recommender agents \cite{lv2026costeercollaborativedecodingtimepersonalization,lv2026specsteersynergizinglocalcontext,zhang2026paradigmusercentricagentplatformcentric,shen2024exploringuserretrievalintegration,yu2025thoughtaugmentedplanningllmpoweredinteractive,wang2025generativelargerecommendationmodels}. These scenarios likewise require balancing evidence sufficiency, noise control, and inference efficiency, which aligns well with our empirical finding that adaptive retrieval serves different roles for weaker and stronger models.

\subsection{Future Evaluation and Application Perspectives}
Our findings also suggest that future work may benefit from evaluating adaptive evidence selection under broader interleaved, agentic, or recommendation-oriented settings, where the boundary between retrieval, ranking, and generation becomes increasingly blurred \cite{zhang2025ragigbenchinnovativeevaluationragbased,zhang2025killingbirdsstoneunifying,yu2025thoughtaugmentedplanningllmpoweredinteractive,zhang2026thinkinghurtsdiagnosingrectifying,wang2025enhancingctrpredictiondecorrelated,Wang_2025,wang2025universalframeworkcompressingembeddings}.


\end{document}